\documentclass[aps,preprint]{revtex4}%
\usepackage{amsfonts}
\usepackage{amsmath}
\usepackage{amssymb}
\usepackage{graphicx}
\usepackage{color}%
\providecommand{\U}[1]{\protect\rule{.1in}{.1in}}

\begin{document}
\preprint{Manuscript }
\title{{\large \textbf{Envelope excitations in electronegative plasmas with electrons
featuring the Tsallis distribution}}}
\author{A. S. Bains}
\email{bainsphysics@yahoo.co.in.}
\affiliation{Institute of Space Sciences, Shandong University at Weihai, 264209 China}
\author{Bo Li}
\email{bbl@sdu.edu.cn}
\affiliation{Institute of Space Sciences, Shandong University at Weihai, 264209 China}
\author{Mouloud Tribeche}
\email{mtribeche@usthb.dz}
\affiliation{Plasma Physics Group (PPG), Theoretical Physics Laboratory (TPL), Faculty of
Physics, University of Bab-Ezzouar, U.S.T.H.B, B.P. 32, El Alia, Algiers
16111, Algeria.}
\keywords{Nonlinear Schr\"{o}dinger equation, Modulational instability, Nonextensivity,
Tsallis statistical mechanics.}
\pacs{52.35.Fp; 52.27.Ep.}

\begin{abstract}
We examine the modulational instability (MI) of ion-acoustic waves (IAWs) in an electronegative plasma containing positive and negative ions as well as electrons that follow the nonextensive statistics proposed by Tsallis [J. Stat. Phys. 52, 479 (1988)]. Using the reductive perturbation method (RPM), the nonlinear Schr\"{o}dinger equation (NLSE) that governs the modulational instability of the IAWs is obtained. Inspired by the experimental work of Ichiki \emph{et al.} [Phys. Plasmas 8, 4275 (2001)], three types of electronegative plasmas are investigated. The effects of various parameters on the propagation of IAWs are discussed in detail numerically. We find that the plasma supports both bright and dark solutions. The presence of the non-extensively distributed electrons is found to play a crucial role in the formation of envelope excitations. The region in the parameter space where the MI exists depends sensitively on the positive to negative ion mass ratio (M) and negative to positive ion density ratio ($\nu$). An extensive range of the nonextensive $q$-parameters {$(-1<q<3)$} is considered and in each case the MI sets in under different conditions. The finding of this investigation is useful for understanding stable wave propagation of envelope ion-acoustic solitary waves in space and laboratory plasmas comprising ions with both positive and negative charges as well as non-Maxwellian electrons.

\end{abstract}
\date{\today}
\startpage{1}
\endpage{102}
\maketitle

\section{Introduction}

Electronegative plasmas, namely, plasmas comprising an appreciable amount of
negative ions, have been found in a wide variety of astrophysical
environments. These include the Earth's ionosphere~\cite{massey} and
mesosphere~\cite{del}, cometary comae~\cite{chaizy}, and the Titan's
ionosphere~\cite{coates}, to name but a few. Equally important is that
electronegative plasmas exhibit many technological applications such as
neutral beam sources \cite{bascal}, plasma processing reactors \cite{gottscho}%
, synthesis of nanomaterials \cite{kokura}, and semiconductor materials
processing \cite{sugai}. Due to its wide application in laboratory and space
plasmas, many authors studied wave propagation phenomena in such plasma
systems \cite{gill, elwaskil, ghosh1, akhter,taibany11, taibany11a}.

There are many circumstances in which the well-known Maxwellian distribution
proves not to be a proper description of the plasma species \cite{christon,
maksimovic, leubner04}. Rather, the so-called superthermal or Tsalis
distribution, seems common to the species in general
\cite{sainia,parvin,tantawy,borhanian}, and the electrons in particular
\cite{pakzad, parvina, shana,ghosha}. As is well-known, the Maxwellian
distribution in the Boltzmann-Gibbs statistics is believed to be universally
valid for macroscopic ergodic equilibrium systems. However, for such systems
as plasmas and gravitational systems where long-range interactions are present and
where non-equilibrium stationary states exist, the Maxwellian distribution may
be inadequate. For this reason, a large number of theoretical investigations
have been made into the nonextensive statistic mechanics based on the
deviations from the Boltzmann--Gibbs--Shannon (B-G-S) entropic measure. A suitable
nonextensive generalization of the B-G-S entropy for statistical equilibrium
was first recognized by Renyi \cite{renyi} and subsequently proposed by
Tsallis\cite{tsallis}, suitably extending the standard additivity of entropy
to the nonlinear, nonextensive case where one particular parameter, the
entropic index $q$, characterizes the degree of nonextensivity of the
considered system. When $q=1$, one recovers the standard, extensive, B-G-S
statistics. This non-additive entropy of Tsallis and the ensuing generalized
statistics have been employed with success in a wide range of phenomena
characterized by nonextensivity \cite{rossignoli,lima1, milovanov,lima2,
abe,kaniadakis, lavagno, wada, reynolds, du, sattin, munoz, wu,leubner1,
hanel, amour}. Specifically, a growing body of observational evidence in
astrophysical environments suggests that the $q$-entropy provides a convenient
framework for the analysis of stellar polytropes \cite{Ref1}, the solar
neutrino problem \cite{Ref2}, and peculiar velocity distribution of galaxy
clusters \cite{Ref3}. Likewise, there are experimental results for
electrostatic plane-wave propagation in a collisionless thermal plasma that
point to a class of Tsallis's velocity distribution described by a
nonextensive $q$-parameter smaller than unity\cite{lima2}.

The deviation of the velocity distribution functions (VDFs) of plasma species
from a Maxwellian has proven to have a profound effect on the nonlinear
collective phenomena in many environments\cite{summer}. It is then no surprise
that the $q$-nonextensive description is becoming increasingly popular as far
as the studies of solitary waves in plasmas where the electron VDF is not a
Maxwellian are concerned\cite{tribeche10,tribeche10a,ali13}. For instance, Saini \emph{et
al.}\cite{saini09} studied Ion-Acoustic Waves (IAWs) with arbitrary amplitudes
in the presence of superthermal electrons and obtained existence conditions
for and some characteristics of {IAWs} in a plasma composed of cold ions and
kappa-distributed electrons. Likewise, Boubakour \textit{et al.}%
\cite{boubakour09} examined such \emph{IAWs} in a plasma comprising inertial
ions, Maxwellian positrons and superthermal electrons, and showed that the
superthermality of the electrons influences substantially the profile of the
solitons. Recently, Tribeche \textit{et al.} \cite{tribeche10} studied the
characteristics of ion-acoustic solitary waves in a two-component plasma with
a q-nonextensive electron velocity distribution. Also, modulational
instability of different modes in several plasma environments has been studied
\cite{gill10,bedi,bains11,bains11b,bains11c}.

Surprisingly, most of the above-mentioned papers are restricted to two species
plasmas, and only a few investigations are concerned with electronegative
plasmas. Among them, Elwaskil \emph{et al.}\cite{elwaskil} studied the
envelope excitations of solitary waves in a system with the combination of
(H$^{+}$,O$_{2}^{-}$) or (H$^{+}$,H$^{-}$) together with nonthermal electrons
whose distribution is described by the Cairns scheme~\cite{cairns}. The model
supported both bright-(pulses) and dark-(holes, voids) solitons. Very
recently, Taibany and Tribeche \cite{taibany12} studied weak solitary
structures in electronegative plasma with nonextensive velocity distribution.
They derived the Korteweg-de Vries (KdV) equation which describes the
evolution of an unmodulated wave, a bare pulse containing no high frequency
oscillation inside the wave packet. It was observed that the nonextensivity of
electrons and the positive to negative ion mass ratio significantly change the
velocity, amplitude and width of the solitary waves.

{The aim of the present study is to extend the work of Taibany and Tribeche
\cite{taibany12}, and examine the combined effect of nonextensively
distributed electrons and negative ions on the modulational instability (MI)
of ion-acoustic waves. For this we derive the nonlinear Schrodinger
equation~(NLSE) that describes the evolution of the modulated wave packet. In
the NLSE, the nonlinearity is in balance with the group velocity dispersion
and the resulting stationary solutions have envelope structures.} The
organization of this paper is as follows. In section II, we present the basic
equations for an electronegative plasma and carry out a reductive perturbation to
derive the appropriate nonlinear Schr\"{o}dinger equation (NLSE). The
modulational instability is discussed in Sec.III. and our findings are
summarized in Sec.IV.

\section{Basic equations}

We consider a plasma model whose constituents are singly charged fluid cold
positive ions (subscript p), singly charged fluid cold negative ions
(subscript n), and nonextensive electrons (subscript e). The nonlinear
dynamics of ion-acoustic oscillations is governed by the following normalized
equations \cite{taibany12}:\newline-for positive ions%

\begin{equation}
\frac{\partial n_{p}}{\partial t}+\frac{\partial n_{p}v_{p}}{\partial x}=0,
\label{1}%
\end{equation}%
\begin{equation}
\frac{\partial u_{p}}{\partial t}+v_{p}\frac{\partial v_{p}}{\partial x}%
+\frac{\partial\phi}{\partial x}=0, \label{2}%
\end{equation}
-for negative ions
\begin{equation}
\frac{\partial n_{n}}{\partial t}+\frac{\partial n_{n}v_{n}}{\partial x}=0,
\label{3}%
\end{equation}%
\begin{equation}
\frac{\partial u_{n}}{\partial t}+v_{n}\frac{\partial v_{n}}{\partial
x}-M\frac{\partial\phi}{\partial x}=0, \label{4}%
\end{equation}
-Poisson equation
\begin{equation}
\frac{\partial^{2}\phi}{\partial x^{2}}=n_{n}-n_{p}+n_{e} \label{5}%
\end{equation}
where the expression for the nonextensive electron density is given by
\cite{bains11,tribeche10a}
\begin{equation}
n_{e}=\mu\lbrack1+(q-1)\phi]^{(q+1)/2(q-1)} \label{6}%
\end{equation}
In Eqs. (\ref{1}) - (\ref{6}), $n_{p,n}$ is the ion number density normalized
by the unperturbed positive ion density $n_{p0}$, $u_{p, n}$ is the ion fluid
velocity normalized by the ion sound speed $C_{s}=(k_{B}T_{e}/m_{p})^{1/2}$,
and $\phi$ is the electrostatic wave potential normalized by $k_{B}T_{e}/e$.
The time $t$ is normalized by the ion plasma period $\omega_{pi}^{-1}=(4\pi
e^{2}n_{p0}/m_{i})^{-1/2}$, and the space variable $x$ is in unit of the ion
Deybe length $\lambda_{Di} =(k_{B}T_{e}/4\pi e^{2}n_{p0})^{1/2}$. Here $k_{B}$
is the Boltzmann constant, $T_{e}$ is the electron temperature, and $e$ is the
electron charge. We define the mass ratio $M$ as $m_{p}/m_{n}$ where $m_{p}$
($m_{n}$) is the positive (negative) ion mass. The neutrality condition
implies $\nu+\mu=1$, where $\nu=n_{n0}/n_{p0}$ and $\mu=n_{e0}/n_{p0}$.
Moreover, the parameter $q$ stands for the strength of the electron
nonextensivity. For $q<-1$, the nonextensive electron distribution is
unnormalizable. In the extensive limiting case $q \rightarrow1$, the electron
density, Eq. (\ref{6}), reduces to its well-known Maxwell-Boltzmann
counterpart.\newline

To investigate the modulation of the IAWs, we employ the standard reductive
perturbation technique (RPT) to derive the appropriate nonlinear
Schr\"{o}dinger equation (NLSE). The independent variables are stretched as
$\xi=\epsilon(x-v_{g}t)$ and $\tau=\epsilon^{2}t$, where $\epsilon$ is a small
parameter and $v_{g}$ the group velocity of the wave. The dependent variables
are then expanded as%
\begin{align}
n_{p}  &  =1+\sum_{m=1}^{\infty}\epsilon^{m}\sum_{l=-\infty}^{+\infty}%
n_{pl}^{m}(\xi,\tau)e^{\iota l(kx-\omega t)},\nonumber\\
n_{n}  &  =\nu+\sum_{m=1}^{\infty}\epsilon^{m}\sum_{l=-\infty}^{+\infty}%
n_{nl}^{m}(\xi,\tau)e^{\iota l(kx-\omega t)},\nonumber\\
u_{p,n}  &  =\sum_{m=1}^{\infty}\epsilon^{m}\sum_{l=-\infty}^{+\infty}%
u_{p,nl}^{m}(\xi,\tau)e^{\iota l(kx-\omega t)},\nonumber\\
\ \phi &  =\sum_{m=1}^{\infty}\epsilon^{m}\sum_{l=-\infty}^{+\infty}\phi
_{l}^{m}(\xi,\tau)e^{\iota l(kx-\omega t)} . \label{7}%
\end{align}
For $n_{p,n}$, $u_{p,n}$ and $\phi$ to be real, one requires that $A_{-l}%
^{m}=\left(  A_{l}^{m}\right)  ^{\ast}$ where the asterisk denotes complex
conjugate. Substituting these expressions along with the stretching
coordinates into Eqs. (\ref{1})-(\ref{5}) and collecting the terms in the
different powers of $\epsilon$, we can obtain the $m$-th order reduced
equations. At the first order $(m=1)$, we can obtain for $l=1$ the first-order
quantities in terms of $\phi_{1}^{(1)}$ as
\begin{align}
-\iota\omega n_{p1}^{(1)}+\iota ku_{p1}^{(1)}=0,\quad-\iota\omega u_{p1}%
^{(1)}+\iota k\phi_{1}^{(1)}  &  =0,\nonumber\\
-\iota\omega n_{n1}^{(1)}+\iota\nu ku_{p1}^{(1)}=0,\quad-\iota\omega
u_{n1}^{(1)}-\iota kM\phi_{1}^{(1)}  &  =0,\nonumber\\
n_{p1}^{(1)}-n_{n1}^{(1)}-(k^{2}+c_{1})\phi_{1}^{(1)}  &  =0,
\end{align}
where $c_{1}=\mu(q+1)/2$. \newline The solution for the first harmonics is
\begin{equation}
n_{p1}^{(1)}=\frac{k^{2}}{\omega^{2}}\phi_{1}^{(1)},\hspace{5mm}u_{p1}%
^{(1)}=\frac{k}{\omega}\phi_{1}^{(1)},\hspace{5mm}n_{n1}^{(1)}=-\nu
M\frac{k^{2}}{\omega^{2}}\phi_{1}^{(1)},\hspace{5mm}u_{n1}^{(1)}=-M\frac
{k}{\omega}\phi_{1}^{(1)}%
\end{equation}
Thus, we obtain the following dispersion relation for IAWs
\begin{equation}
\frac{\omega^{2}}{k^{2}}=\frac{1+\nu M}{(k^{2}+c_{1})}.
\end{equation}

\begin{figure}[ptb]
\includegraphics[width=10cm]{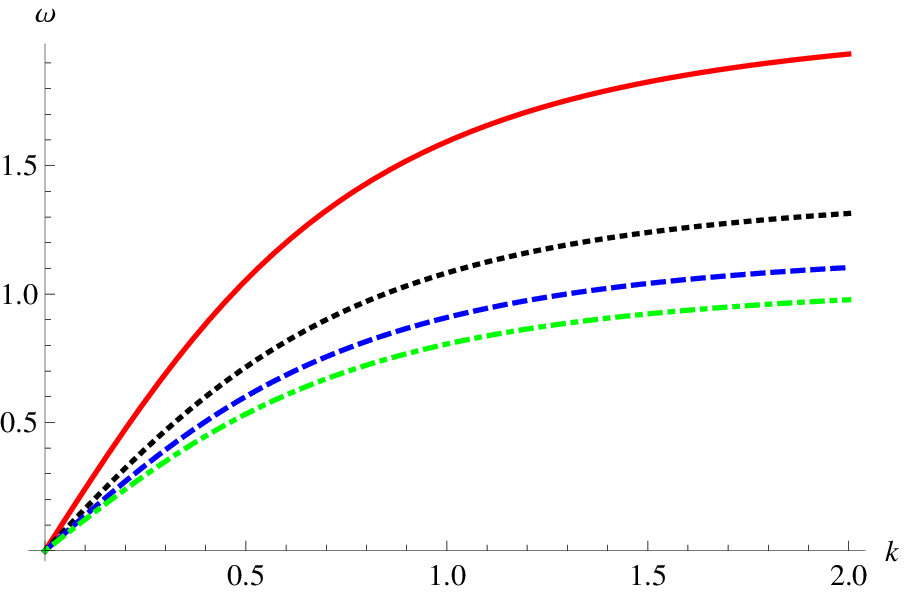}\newline\caption{(Color online) Variation of
the carrier frequency $\omega$ with the wave number $k$ for different types of
plasmas with $\mu=0.5$ and $q= 0.5$. Solid curve corresponds to $Xe^{+} -
F^{-}$ plasma $(M = \frac{131}{19} \simeq6.89)$; dotted curve to $Ar^{+} -
F^{-}$ plasma $(M= \frac{40}{19} \simeq2.10)$, dashed curve to $Xe^{+} -
SF_{6}^{-}$ plasma $(M= \frac{131}{146} \simeq0.89)$ and dotted dashed curve
to $Ar^{+} - SF_{6}^{-}$ plasma $(M= \frac{40}{146} \simeq0.27)$.}%
\label{fig1}%
\end{figure}

Figure~\ref{fig1} presents the solution to the dispersion relation in the form
of $\omega$ as a function of $k$ at different values of the positive to
negative mass ratio $M$. It is observed that $\omega$ increases with
increasing $k$. Furthermore, $\omega$ decreases with increasing $M$. This
means that increasing $M$ decreases the energy as well as the oscillations of
the wave.

At the second order $(m=2)$ and for $l=1$, the following equations are
obtained
\begin{equation}
-\iota\omega n_{p1}^{(2)}+\iota ku_{p1}^{(2)}=v_{g}\frac{\partial n_{p1}%
^{(1)}}{\partial\xi}-\frac{\partial u_{p1}^{(1)}}{\partial\xi},
\end{equation}%
\begin{equation}
-\iota\omega u_{p1}^{(2)}+\iota k\phi_{1}^{(2)}=v_{g}\frac{\partial
u_{p1}^{(1)}}{\partial\xi}-\frac{\partial\phi_{1}^{(1)}}{\partial\xi},
\end{equation}

\begin{equation}
-\iota\omega n_{n1}^{(2)}+\iota k \nu u_{n1}^{(2)}= v_{g}\frac{\partial
n_{n1}^{(1)}}{\partial\xi}- \nu\frac{\partial u_{n1}^{(1)}}{\partial\xi},
\end{equation}%
\begin{equation}
-\iota\omega u_{n1}^{(2)}-\iota k M \phi_{1}^{(2)}= v_{g}\frac{\partial
u_{n1}^{(1)}}{\partial\xi}+ M\frac{\partial\phi_{1}^{(1)}}{\partial\xi},
\end{equation}

\begin{equation}
(k^{2}+c_{1})\phi_{1}^{(2)}+n_{n1}^{(2)} -n_{p1}^{(2)} =2\iota k\frac
{\partial\phi_{1}^{(1)}}{\partial\xi},
\end{equation}
along with the compatibility condition
\begin{equation}
v_{g}=\frac{c_{1}}{(1+ \nu M)}\frac{\omega^{3}}{k^{3}}.
\end{equation}
If we proceed further in the same way as in Ref. \cite{kourakis}, and
substitute the derived expressions for $m=2$, $l=0$ and $m=2$, $l=2$ into the
components for $m=3$, $l=1$ of the reduced equations, we can obtain the
following nonlinear Schrodinger equation%

\begin{equation}
\iota\frac{\partial\phi_{1}^{(1)}}{\partial\tau}+P\frac{\partial^{2}\phi
_{1}^{(1)}}{\partial\xi^{2}}+Q|\phi_{1}^{(1)}|^{2}\phi_{1}^{(1)}=0
\label{nlse}%
\end{equation}
where%

\begin{equation}
P=\frac{1}{2}\frac{d^{2}\omega}{dk^{2}}=-\frac{3}{2}\frac{c_{1}}{(1+\nu
M)^{2}}\frac{\omega^{5}}{k^{4}}%
\end{equation}
and
\begin{align}
Q  &  =\frac{\omega^{3}}{2k^{2}}\left[  3c_{3}-2c_{2}(A_{\phi}+B_{\phi
})-2\frac{k^{3}}{\omega^{3}}(A_{up}+B_{up})-\frac{k^{2}}{\omega^{2}}%
(A_{np}+B_{np})\right. \nonumber\\
&  \left.  -2\nu M\frac{k^{3}}{\omega^{3}}(A_{un}+B_{un})-M\frac{k^{2}}%
{\omega^{2}}(A_{nn}+B_{nn})\right]
\end{align}
where
\begin{align}
A_{\phi}  &  =\frac{(1-\nu M^{2})\frac{3k^{4}}{2\omega^{4}}+c_{2}}%
{(c_{1}+4k^{2})-(1-\nu M)\frac{k^{2}}{\omega^{2}}},\nonumber\\
B_{\phi}  &  =\frac{v_{g}^{2}\left(  2c_{2}+(1/v_{g}+2k/\omega)\frac
{(1-/nuM^{2})k^{2}}{v_{g}\omega^{2}}\right)  }{v_{g}^{2}c_{1}-(1+\nu
M)},\nonumber\\
A_{up}  &  =A_{\phi}\frac{k}{\omega}+\frac{k^{3}}{2\omega^{3}},\quad
A_{np}=A_{up}\frac{k}{\omega}+\frac{k^{4}}{\omega^{4}},\nonumber\\
A_{un}  &  =-MA_{\phi}\frac{k}{\omega}+\frac{M^{2}k^{3}}{2\omega^{3}},\quad
A_{nn}=A_{un}\frac{\nu k}{\omega}+\frac{\nu M^{2}k^{4}}{\omega^{4}%
},\nonumber\\
B_{up}  &  =\frac{B_{\phi}}{v_{g}}+\frac{k^{2}}{v_{g}\omega^{2}},\quad
B_{np}=\frac{B_{up}}{v_{g}}+\frac{2k^{3}}{v_{g}\omega^{3}}\nonumber\\
B_{un}  &  =-B_{\phi}\frac{M}{v_{g}}+\frac{M^{2}k^{2}}{v_{g}\omega^{2}},\quad
B_{un}=B_{up}\frac{\nu}{v_{g}}+\frac{2\nu M^{2}k^{3}}{v_{g}\omega^{3}%
}\nonumber\\
c_{2}  &  =\frac{(q+1)(q-3)}{8},\quad c_{3}=\frac{(q+1)(q-3)(3q-5)}{48}%
\end{align}

\section{Numerical Analysis}

Let us now investigate the stability/instability of the modulated wave packets
in an electronegative plasma with nonextensive electrons on the basis of the
NLSE (\ref{nlse}) that governs the MI of the IAWs. Based on the linear
stability analysis \cite{amin}, it is observed that the wave is modulationally
unstable when $PQ>0$ in the modulation wave number region $k^{2}<\frac{2Q}%
{P}|\phi_{0}|^{2}$, where $\phi_{0}$ is the amplitude of the carrier waves.
Furthermore, the maximum growth rate is given by $Q|\phi_{0}|^{2}$ and is
attained at $k=\sqrt{\frac{Q}{P}}|\phi_{0}|$. Two types of stationary
solutions are possible: (i) bright envelope soliton when $PQ>0$ and (ii) dark
envelope soliton when $PQ<0$. For unstable wave packets ($PQ>0$), we have
envelope solitons given by
\begin{equation}
\phi=\sqrt{\left|  \frac{2\gamma}{Q}\right|  }\mathrm{sech}\left(  \left|
\frac{\gamma}{P}\right|  \zeta\right)  e^{\iota\gamma\tau}, \label{PQ}%
\end{equation}
where $\gamma$ is a real constant. For stable wave packet ($PQ<0$), we obtain
a modulationaly stable ion-acoustic mode with special solution known as
envelope dark soliton given by
\begin{equation}
\phi=\sqrt{\left|  \frac{\gamma}{Q}\right|  }\tanh\left(  \left|  \frac
{\gamma}{2P}\right|  \zeta\right)  e^{\iota\gamma\tau}. \label{QP}%
\end{equation}
It is obvious from Eqs.(\ref{PQ}) and (\ref{QP}) that the width and the
amplitude of the solitary wave vary with $P$ and $Q$. The soliton width is
proportional to $|P|$ and the soliton amplitude is inversely proportional to
$|Q|$. These coefficients in turn depend upon a number of parameters such as
positive to negative ion mass ratio $M$, negative to positive ion density
ratio $\nu$ and the electron nonextensive parameter $q$. Consequently, one
expects that these parameters would affect the stability criteria over a wide
range in the parameters space.

To proceed further, some specific applications are necessary for a
quantitative analysis to be made. To this end, we focus on the following four
types of plasmas~{\cite{ichiki}}: (Xe$^{+}$-F$^{-}$) with $M={131}/{19}%
\simeq 6.89$, (Ar$^{+}$-F$^{-}$) with $M={40}/{19}\simeq 2.10$, (Xe$^{+}%
$-SF$_{6}^{-}$) with $M={131}/{146}\simeq 0.89$ and (Ar$^{+}$-SF$_{6}^{-}$)
with $M= {40}/{146}\simeq 0.27$. Evaluating the dispersive coefficient $P$, we
find that for all sets of parameters, $P$ is always negative. Hence, the
conditions for instability is determined by the sign of the nonlinear
coefficient $Q$.

\begin{figure}[ptb]
\includegraphics[width=10cm]{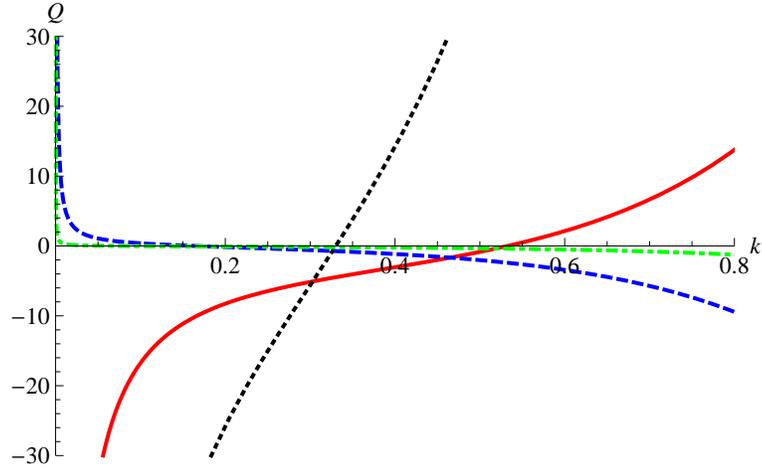}\newline\caption{(Color online) Variation of
the NLSE coefficient $Q$ with the carrier wave number $k$ for different types
of plasmas with $\mu=0.5$ and $q= 0.5$. Solid curve corresponds to $Xe^{+} -
F^{-}$ plasma $(M= \frac{131}{19} \simeq6.89)$; dotted curve to $Ar^{+} -
F^{-}$ plasma $(M= \frac{40}{19} \simeq2.10)$, dashed curve to $Xe^{+} -
SF_{6}^{-}$ plasma $(M = \frac{131}{146} \simeq0.89)$ and dotted dashed curve
to $Ar^{+} - SF_{6}^{-}$ plasma $(M = \frac{40}{146} \simeq0.27)$.}%
\label{fig2}%
\end{figure}

Figure~\ref{fig2} displays the variation of the nonlinear coefficient $Q$ for
different types of plasmas. It is clear from this plot that $Q$ has different
trends for the different types of plasmas. For higher values of positive to
negative ion mass ratio $M$, the nonlinear coefficient is negative for small
$k$ and becomes positive for larger values of $k$. As $M$ decreases, $Q$
starts to change its sign towards positive values for small wave numbers. The
behavior is totally different when $M<1$, for the Xe$^{+}$-SF$_{6}^{-}$ and
Ar$^{+}$-SF$_{6}^{-}$ plasmas. In this case the nonlinear coefficient $Q$ is
positive for small values of $k$, but turns negative beyond a certain $k$.

Given that $P$ is negative definite for all the parameters considered, the
product $PQ$ has different trends for different types of plasmas as well. It
follows that the stability profile may be readily obtained by examining the
ratio $P/Q$ against $k$. The wavenumber $k$ corresponding to $Q=0$ is of
particular relevance since it is the critical or threshold wavenumber for the
onset of the MI. For the ease of description, it will be termed $k_{c}$.

\begin{figure}[ptb]
\includegraphics[width=10cm]{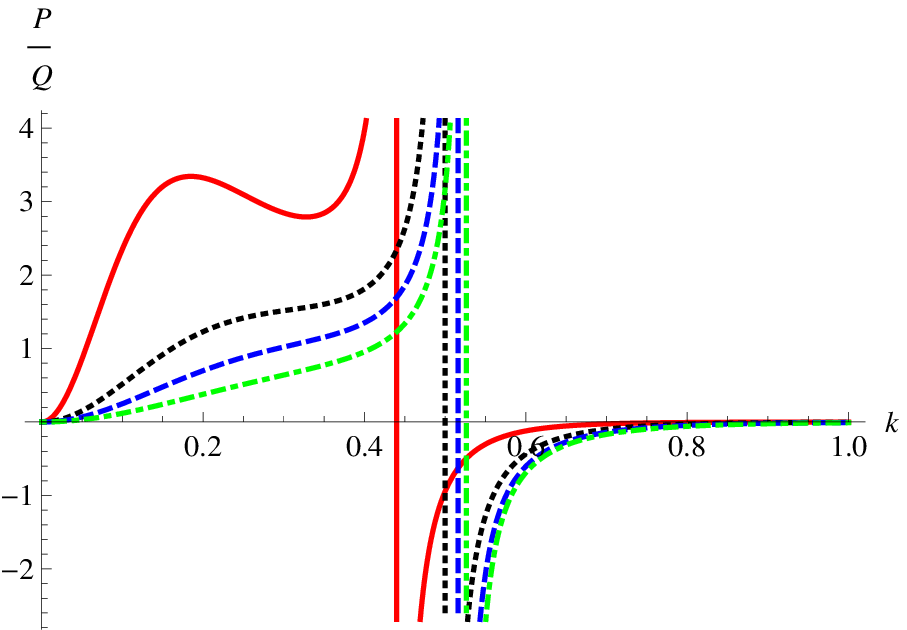}\newline\caption{(Color online) Variation of
the ratio $P/Q$ with the carrier wave number $k$ for different values of the
$q$-nonextensive parameter with $\mu=0.5$ for $Xe^{+} - F^{-}$ plasma $(M =
\frac{131}{19} \simeq6.89)$. Solid curve corresponds to $q=-0.7$; dotted curve
to $q=-0.3$; dashed curve to $q=0$ and dotted-dashed curve to $q=0.4$.}%
\label{fig3}%
\end{figure}

For the Xe$^{+}$-F$^{-}$ plasma, the variation of $P/Q$ as a function of $k$
for different values of the nonextensive parameter $q$ is shown in
Fig.\ref{fig3}. One notices that both unstable and stable regions are
obtained. The amplitude remains unstable when $k<k_{c}$ and becomes stable
when the opposite is true. Bright solitons occur in the former case and
therefore correspond to large wavelengths, whereas dark envelope solitons
occur in the latter region. Besides, it can be seen that the critical value
$k_{c}$, where the instability sets in, increases when $q$ increases, meaning
that for vanishingly small $q \rightarrow1 $, the MI occurs at small wavelengths.

\begin{figure}[ptb]
\includegraphics[width=10cm]{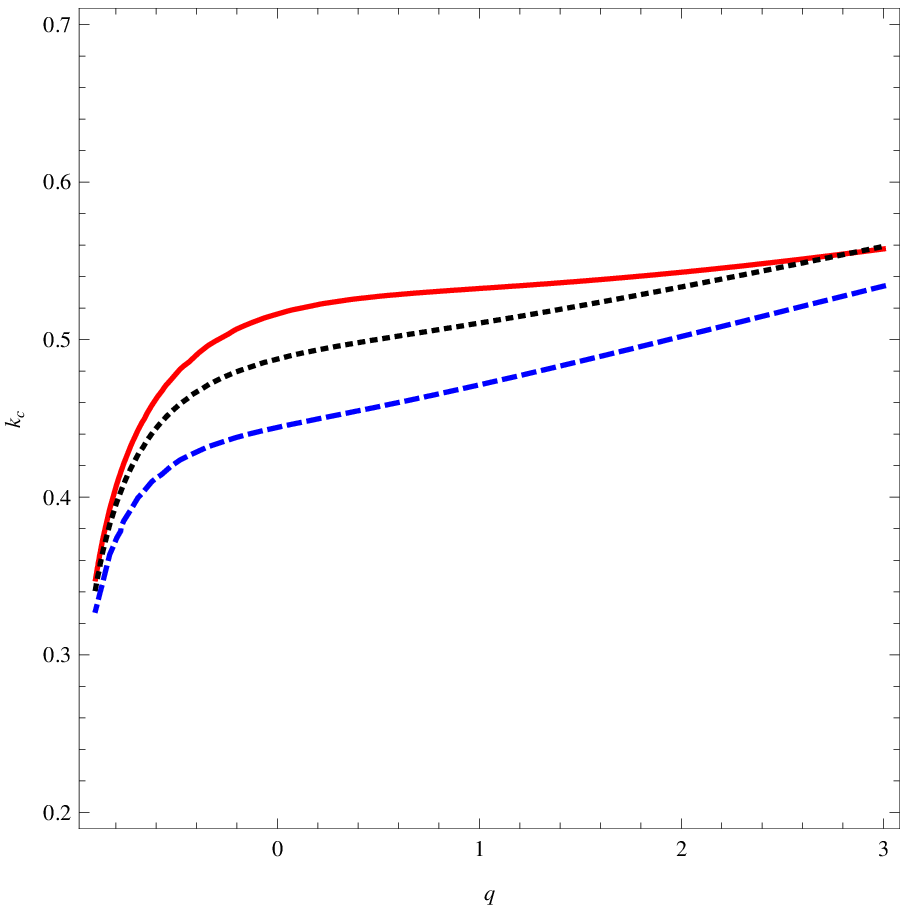}\newline\caption{(Color online) Variation of
the critical wave number $k_{c}$ with the $q$-nonextensive parameter for
different values of $\mu$ with $M = 6.89$. Solid curve corresponds to
$\mu=0.5$; dotted curve to $\mu=0.6$ and dashed curve to $\mu=0.7$.}%
\label{fig4}%
\end{figure}

The combined effect of negative to positive ion ratio ($\nu$) and the electron
nonextensivity for $M=6.89$ is depicted in Fig.\ref{fig4}. It is clear that
the critical wavenumber $k_{c}$ decreases with increasing $\mu$, namely, with
the decrease in the negative to positive ion density ratio. In other words,
the more negative ions that are introduced into the plasma, the smaller the
wavelength beyond which the instability sets in will be. For a fixed value of
$\mu$, the critical wavenumber $k_{c}$ increases sharply in the region
$-0.9<q<0$. As $q\rightarrow1$, i.e., as the distribution approaches a
Maxwellian, $k_{c}$ acquires an almost saturated value.

\begin{figure}[ptb]
\includegraphics[width=10cm]{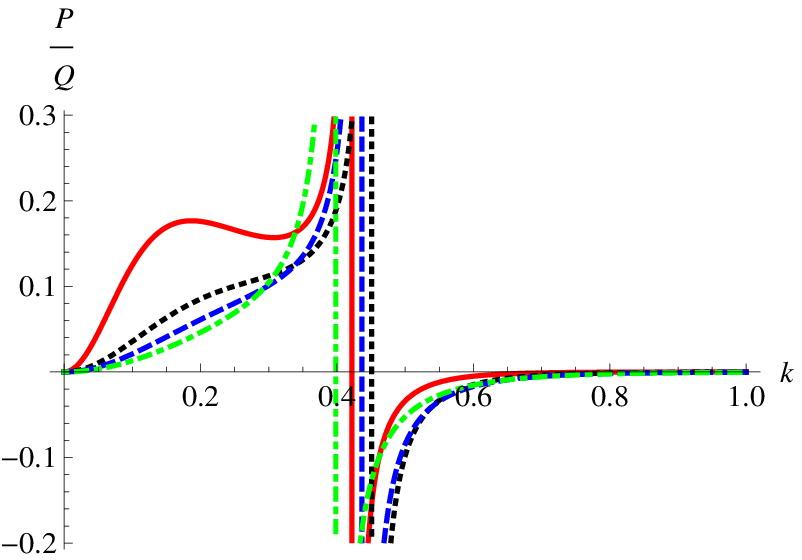}\newline\caption{(Color online) Variation of
the ratio $P/Q$ with the carrier wave number $k$ for different values of the
$q$-nonextensive parameter with $\mu=0.5$ for the $Ar^{+} - F^{-}$ plasma $(M
= \frac{40}{19} \simeq2.10)$. Solid curve corresponds to $q=-0.7$; dotted
curve to $q=-0.3$; dashed curve to $q=0$ and dotted-dashed curve to $q=0.4$.}%
\label{fig5}%
\end{figure}

With $\mu$ fixed, the variation of $P/Q$ as a function of $k$ for an Ar$^{+}%
$-F$^{-}$ plasma $(M=\frac{40}{19}\simeq2.10)$ is shown in Fig.\ref{fig5}.
Qualitatively similar behaviors are obtained as earlier, and both unstable and
stable regions are obtained. The amplitude is unstable at large wavelengths
and becomes stable as the wavelength drops below a certain value. The electron
nonextensivity has an interesting effect on the instability profile now. The
critical wavenumber $k_{c}$ increases for negative values of $q$. However, a
further increase in the $q-$parameter decreases $k_{c}$. As the distribution
approaches a Maxwellian the stability/instability occurs at small $k$.

\begin{figure}[ptb]
\includegraphics[width=10cm]{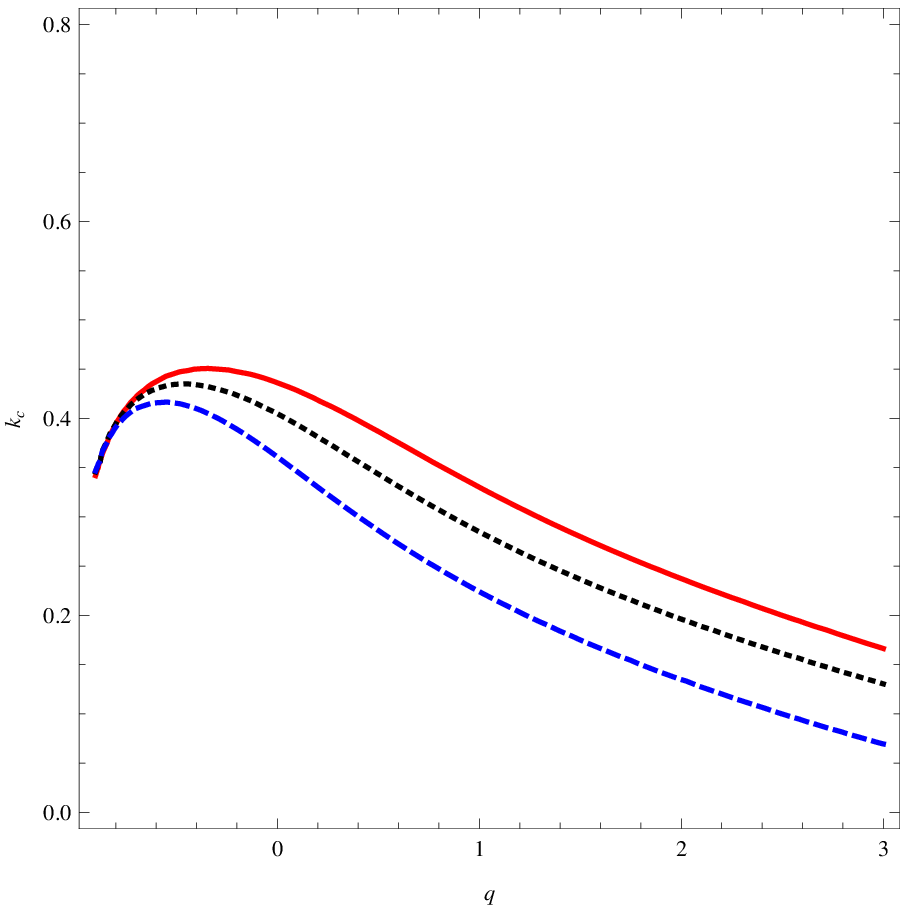}\newline\caption{(Color online) Variation of
the critical wave number $k_{c}$ with the $q$-nonextensive parameter for
different values of $\mu$ with $M = 2.10$ . Solid curve corresponds to
$\mu=0.5$; dotted curve to $\mu=0.6$ and dashed curve to $\mu=0.7$.}%
\label{fig6}%
\end{figure}

This picture is clearer in Fig.\ref{fig6}, where the critical wavenumber
$k_{c}$ is plotted as a function of $q$ for different values of $\mu$. At any
given value of $q$, it is seen that with an increase in $\mu$, the critical
value decreases, just like in the previous case. The amplitude is unstable at
small $k$, and the critical wavenumber at which the transition occurs
increases in the region $-0.9<q<-0.5$. After that, the critical number
continuously decreases with increasing $q$.

In the two plasmas we have discussed so far, the amplitude is unstable at
small $k$ (large wavelength) and becomes stable at large $k$. The results are
very different from earlier investigations, due to large values of $M$. The
available studies have not examined the case where $M>1$. {The maximum value
of $M=1$ is taken by Elwaskil \emph{et al. }\cite{elwaskil} for a H$^{+}%
$-H$^{-}$ plasma. So it is worth mentioning here that $M$ has significant
effect on the formation of envelope solitons, \emph{i.e.}, if we take $M>1$
the results are opposite to the case for $M<1$}.

\begin{figure}[ptb]
\includegraphics[width=10cm]{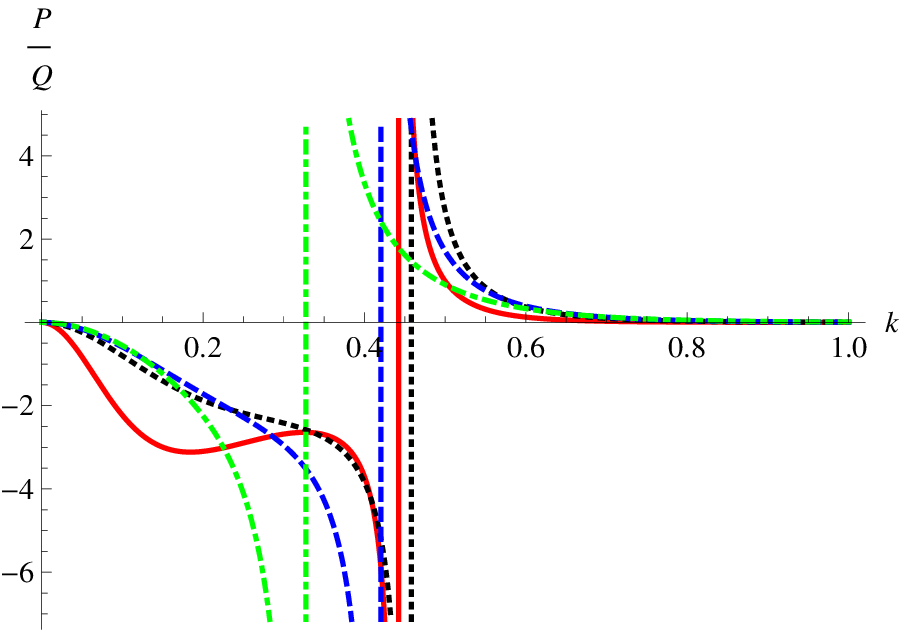}\newline\caption{(Color online) Variation of
the ratio $P/Q$ with the carrier wave number $k$ for different values of the
$q$-nonextensive parameter with $\mu=0.5$ for $Xe^{+} - SF_{6}^{-}$ plasma $(M
= \frac{131}{146} \simeq0.89)$. Solid curve corresponds to $q=-0.7$; dotted
curve to $q=-0.3$; dashed curve to $q=0$ and dotted-dashed curve to $q=0.4$.}%
\label{fig7}%
\end{figure}

\begin{figure}[ptb]
\includegraphics[width=10cm]{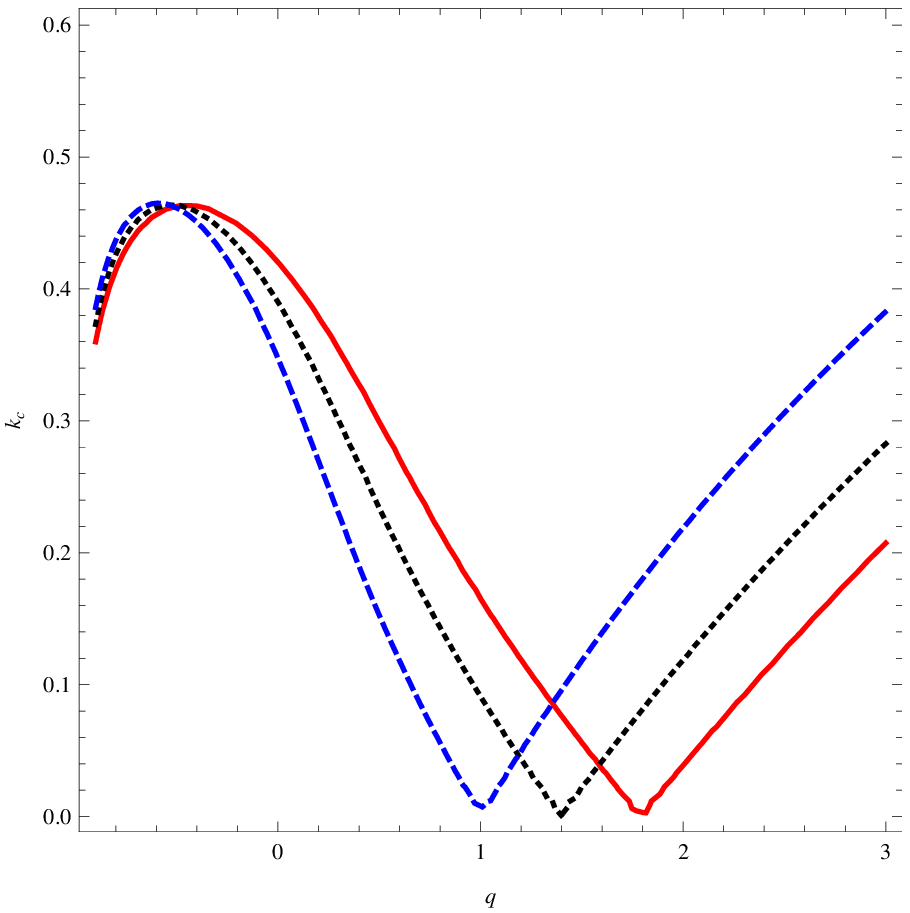}\newline\caption{(Color online) Variation of
the critical wave number $k_{c}$ with the $q$-nonextensive parameter for
different values of $\mu$ with $M = 0.89$ . Solid curve corresponds to
$\mu=0.5$; dotted curve to $\mu=0.6$ and dashed curve to $\mu=0.7$.}%
\label{fig8}%
\end{figure}

Now we proceed to the Xe$^{+}$-SF$_{6}^{-}$ and Ar$^{+}$-SF$_{6}^{-}$ plasmas,
where $M<1$. It turns out that the results are in many ways distinct from what
we found. The effect of the electron nonextensivity on the modulational
instability in the case of the Xe$^{+}$-SF$_{6}^{-}$ plasma is studied in
detail in Figs.\ref{fig7} and \ref{fig8}. The former, which depicts the
ratio~${P/Q}$ as a function of $k$ for different values of $q$, at a given
value of $\mu$ displays such an influence. Although both dark and bright
excitations are possible, as in the previous cases, what is distinct now is
that the wave is stable for small $k$, but modulationally unstable at large
$k$. The critical wavenumber $k_{c}$ at which the instability sets in first
increases and then decreases for different ranges of $q$. The electron
nonextensivity plays therefore a crucial role in the characterization of
localized envelope excitations in different plasma environments. This point is
better illustrated in Fig.\ref{fig8}, in which a wide range of $q$ together
with a number of $\mu$ are explored. Note that the critical wavenumber $k_{c}$
behaves in different manners in different ranges of the nonextensive parameter
$q$. For $q<-0.5$, the critical wavenumber increases with increasing $\mu$.
However, as $q\rightarrow1$, the critical value decreases with increasing
$\mu$. This means that adding more negative ions into the system leads to
instability at larger ${k}$. Similar results (not shown) are obtained in the
case of the Ar$^{+}$-SF$_{6}^{-}$ plasma for which $M=0.27$. The amplitude is
stable for small wavenumbers and becomes unstable at large ${k}$. The critical
wavenumber for instability first increases and then decreases with an increase
in the ${q}$ parameter.

\section{Conclusion}

To conclude, we have investigated the modulational instability (MI) of
ion-acoustic waves (IAWs) in an electronegative plasma where both positively and
negatively charged ions are present, and where the electrons are described
within the theoretical framework of the nonextensive statistics proposed by
Tsallis \cite{tsallis}. Using the reductive perturbation method (RPM), the
nonlinear Schr\"{o}dinger equation (NLSE) that governs the modulational
instability of the IAWs is obtained. Inspired by the experimental work of
Ichiki \textit{et al.} \cite{ichiki}, four types of electronegative plasmas
are examined, which differ mainly in the positive to negative ion mass ratio.
The effects of various parameters on the propagation of the IAWs are discussed
in detail numerically. It is found that the plasma supports both bright as
well as dark solitons. We find that the presence of the nonextensive electrons
plays a crucial role in the formation of the envelope excitations. The region
of modulational instability is significantly affected by the positive to
negative ion mass ratio ($M$), and the negative to positive ion density ratio
($\nu$). An extensive range of the nonextensive $q$-parameter is considered
and in each case the MI sets in under different conditions. The findings of
this investigation are useful in understanding stable wave propagation of
envelope ion-acoustic solitary waves in space and laboratory plasmas where two
type of ions and non-Maxwellian electrons are present.

\acknowledgements
This research is supported by the National Natural Science Foundation of China
(40904047, 41174154, and 41274176), the Ministry of Education of China
(20110131110058 and NCET-11-0305), and by the Provincial Natural Science
Foundation of Shandong via Grant JQ201212.

\newpage

\begin{thebibliography}{99}                                                                                               %


\bibitem {massey}H. Massey, Negative Ions, 3rd ed. (Cambridge University
Press, Cambridge, 1976), p. 663.

\bibitem {del}C. F. del Pozo, E. Turunen, and T. Ulich,Ann. Geophys., 17,
782-793, (1999)

\bibitem {chaizy}P. H. Chaizy, H. Re`me, J. A. Sauvaud, C. D'uston, R. P. Lin,
D. E. Larson, D. L. Mitchell, K. A. Anderson, C. W. Carlson, A. Korth, and D.
A. Mendis, Nature (London) 349, 393 (1991).

\bibitem {coates}A. J. Coates, F. J. Crary, G. R. Lewis, D. T. Young, J. H.
Waite, Jr., and E. C. Sittler Jr., Geophys. Res. Lett. 34, L22103, (2007).

\bibitem {bascal}M. Bascal and G. W. Hamilton, Phys. Rev. Lett. 42, 1538 (1979).

\bibitem {gottscho}R. A. Gottscho and C. E. Gaebe, IEEE Trans. Plasma Sci. 14,
92 (1986).

\bibitem {kokura}H. Kokura, S. Yoneda, K. Nakamura, N. Mitsuhira, M. Nakamura, and

\bibitem {sugai}H. Sugai, Jpn. J. Appl. Phys. 38(Part 1), 5256 (1999).

\bibitem {gill}T. S. Gill, H. Kaur and N. S. Saini, Phys. Plasmas, 10, 3927 (2003).

\bibitem {elwaskil}S. A. Elwaskil, E. K. El-Shewy and H. G. Abdelwahed, Phys.
Plasmas 17, 052301 (2010).

\bibitem {ghosh1}S. Ghosh, S. Sarkar, M. khan and M. R. Gupta. Phys. Rev E,
84, 066401 (2011).

\bibitem {akhter}T. Akhter, M.M. Hossain and A.A. Mamun, Astrophys. Space Sci.
345, 283 (2013).

\bibitem {taibany11}W. F. El-Taibany, N. A. El-Bedwehy, and E. F. El-Shamy,
Phys. Plasmas18, 033703 (2011).

\bibitem {taibany11a}S. K. El-Labany, W. M. Moslem, Kh. A. Shnishin, S. A.
El-Tantawy, and P. K. Shukla, Phys. Plasmas 18, 042306 (2011).

\bibitem {christon}S. P. Christon, D. G. Mitchell, D. J. Williams, L. A.
Frank, C. Y. Huang, and T. E. Eastman, J. Geophys. Res. 93, 2562, (1988).

\bibitem {maksimovic}M. Maksimovic, V. Pierrard, and P. Riley, Geophys. Res.
Lett. 24, 1511, (1997).

\bibitem {leubner04}M. P. Leubner, Phys. Plasmas 11, 1308 (2004).

\bibitem {sainia}N. S. Saini and I.Kourakis, Phys. Plasmas, 15, 123701 (2008).

\bibitem {parvin}P. Eslami, M. Mottaghizadeh, H. R. Pakzad, Phys. Plasmas 18,
102303 (2011).

\bibitem {tantawy}S. A. El-Tantawy, N. A. El-Bedwehy and W. M. Moslem, Phys.
Plasmas, 18 052113 (2011).

\bibitem {borhanian}J. Borhanian and M. Shahmansouri, Phys. Plasmas 20 013707 (2013)

\bibitem {pakzad}H. R. Pakzad, Phys. Plasmas. 18, 082105 (2011).

\bibitem {parvina}P. Eslami, M. Mottaghizadeh, H. R. Pakzad, Phys. Plasmas 18,
102313 (2011).

\bibitem {shana}S. A. Shan and A. Mushtaq, Phy. Scr. 86, 035503 (2012).

\bibitem {ghosha}D. K. Ghosh, P. Chatterjee and U. G. Narayan, Phys. Plasmas,
19, 033703 (2012).

\bibitem {renyi}A. Renyi, Acta Math. Hungaria \textbf{6}, 285 (1955).

\bibitem {tsallis}C. Tsallis, J. Stat. Phys. \textbf{52}, 479 (1988).

\bibitem {rossignoli}R. Rossignoli and N. Canosa, Phys. Lett. A \textbf{264}%
,148 (1999).

\bibitem {lima1}A. R. Lima, J. S. S\'{a} Martins, and T. J. P. Penna, Physica
A \textbf{268}, 553 (1999).

\bibitem {milovanov}A. V. Milovanov and L. M. Zelenyi, Nonlin. Processes
Geophys. \textbf{7}, 211 (2000).

\bibitem {abe}S. Abe, S. Martinez, F. Pennini, and A. Plastino, Phys. Lett. A
\textbf{281}, 126 (2001).

\bibitem {kaniadakis}G. Kaniadakis, Phys. Lett. A \textbf{288}, 283 (2001).

\bibitem {lavagno}A. Lavagno, Phys. Lett. A \textbf{301}, 13 (2002).

\bibitem {wada}T. Wada, Phys. Lett. A \textbf{297}, 334 (2002).

\bibitem {reynolds}A. M. Reynolds and M. Veneziani, Phys. Lett. A
\textbf{327}, 9 (2004).

\bibitem {du}J. Du, Phys. Lett. A \textbf{320}, 347 (2004).

\bibitem {sattin}F. Sattin, Phys. Scr. \textbf{71}, 443 (2005).

\bibitem {munoz}V. Mu\~{n}oz, Nonlin. Processes Geophys. \textbf{13}, 237 (2006).

\bibitem {wu}J. Wu and H. Chen, Phys. Scr. \textbf{75}, 722 (2007).

\bibitem {leubner1}M. P. Leubner, Nonlin. Processes Geophys. \textbf{15}, 531 (2008).

\bibitem {hanel}R. Hanel and S. Thurner, Phys. Lett. A \textbf{373}, 1415 (2009).

\bibitem {amour}R. Amour and M. Tribeche, Phys. Plasmas \textbf{17}, 063702 (2010).

\bibitem {Ref1}A. R. Plastino and A. Plastino, Phys. Lett. A \textbf{174}, 384 (1993).

\bibitem {Ref2}G. Kaniadakis, A. Lavagno, and P. Quarati, Phys. Lett. B
\textbf{369}, 308 (1996).

\bibitem {Ref3} A. Lavagno, G. Kaniadakis, M. Rego-Monteiro, P. Quarati, and
C. Tsallis, Astrophys. Lett. Commun. \textbf{35}, 449 (1998).

\bibitem {lima2}J. A. S. Lima, R. Silva, Jr., and J. Santos, Phys. Rev. E
\textbf{61}, 3260 (2000).

\bibitem {summer}D. Summers and R. M. Thorne, Phys. Fluids B 3, 1835 (1991).

\bibitem {tribeche10}M. Tribeche, L. Djebarni, R. Amour, Phys. Plasmas
\textbf{17} 042114 (2010).

\bibitem {tribeche10a}M. Tribeche and L. Djebarni, Phys. Plasmas 17, 124502 (2010).

\bibitem {ali13}S. Ali Shan and N. Akhtar, Astrophys. Space Sci. 346, 367 (2013).

\bibitem {saini09}N. S. Saini, I. Kourakis and M. A. Hellberg, Phys. Plasmas.
16, 062903 (2009).

\bibitem {boubakour09}N. Boubakour, M. Tribeche and K. Aoutou, Phys. Scr.
\textbf{79} 065503 (2009).

\bibitem {gill10}T. S. Gill, A. S. Bains and C. Bedi, Phys. Plasmas
\textbf{17}, 013701 (2010).

\bibitem {bedi}T. S. Gill, C. Bedi and A. S. Bains, Phys. Scr. \textbf{81}
055503 (2010).

\bibitem {bains11}A. S. Bains, M. Tribeche and T. S. Gill, Phys. Plasmas, 18,
022108 (2011).

\bibitem {bains11b}A. S. Bains, M. Tribeche and T. S. Gill, Phys. Letters A,
375, 2059 (2011).

\bibitem {bains11c}A. S. Bains, M. Tribeche and C. S. Ng, Astro. Space Sci.
343, 621 (2013).

\bibitem {cairns}R.A. Cairns, A.A. Mamun, R. Bingham, R. Bostr\"{o}m, R.O.
Dendy, C.M.C. Nairn, P.K. Shukla, Geophys. Res. Lett. 22, 2709 (1995).

\bibitem {taibany12}W. F. El Taibany and Tribeche Phys. Plasmas 19, 024507 (2012).

\bibitem {kourakis}I. Kourakis and P. K. Shukla, Nonlinear Process. Geophys.
12, 407 (2005).

\bibitem {amin}M. R. Amin, G. E. Morfill,P. K. Shukla: Phys. Rev. E, 58, 6517 (1998).

\bibitem {ichiki}R. Ichiki, M. Shindo, S. Yoshimura, T. Watanabe, and Y.
Kawai, Phys. Plasmas 8, 4275 (2001).
\end{thebibliography}
\end{document}